\documentclass[structabstract]{aa}

\usepackage{natbib} 
\bibpunct{(}{)}{;}{a}{}{,} 
\usepackage{graphicx}
\usepackage{amsmath}
\usepackage{txfonts}
\usepackage{xspace}
\usepackage{amssymb}
\newcommand{\micron}{\mbox{$\mathrm{\mu m}$}\xspace}
\newcommand{\spitz}{{\it Spitzer}\xspace}
\newcommand{\rhk}{$\log R'_{\rm HK}$\xspace}

\newcommand{\feh}{\mbox{[Fe/H]}}
\newcommand{\teff}{\mbox{$T_{\rm eff}$}}

\newcommand{\vsini}{\mbox{$v \sin i$}}

\newcommand{\kms}{\mbox{km\,s$^{-1}$}}
\newcommand{\ms}{\mbox{m\,s$^{-1}$}}

\newcommand{\mplanet}{\mbox{$M_{\rm pl}$}}
\newcommand{\rplanet}{\mbox{$R_{\rm pl}$}}
\newcommand{\densplanet}{\mbox{$\rho_{\rm pl}$}}
\newcommand{\mjup}{\mbox{$M_{\rm Jup}$}}
\newcommand{\rjup}{\mbox{$R_{\rm Jup}$}}
\newcommand{\densjup}{\mbox{$\rho_{\rm Jup}$}}
\newcommand{\mstar}{\mbox{$M_{*}$}}
\newcommand{\rstar}{\mbox{$R_{*}$}}
\newcommand{\densstar}{\mbox{$\rho_*$}}
\newcommand{\msol}{\mbox{$M_\odot$}}
\newcommand{\rsol}{\mbox{$R_\odot$}}
\newcommand{\denssol}{\mbox{$\rho_\odot$}}
\newcommand{\secos}{\mbox{$\sqrt{e} \cos \omega$}} 
\newcommand{\sesin}{\mbox{$\sqrt{e} \sin \omega$}\xspace} 
\newcommand{\ecos}{\mbox{$e \cos \omega$}} 
\newcommand{\esin}{\mbox{$e \sin \omega$}} 
\newcommand{\svcos}{\mbox{$\sqrt{\vsini} \cos \lambda$}} 
\newcommand{\svsin}{\mbox{$\sqrt{\vsini} \sin \lambda$}}

\newcommand{\iracone}{\mbox{$\Delta F_{3.6}$}} 
\newcommand{\iractwo}{\mbox{$\Delta F_{4.5}$}}

\def\teqlf{$T_{\rm P,A=0,f}$}
\def\teqlfull{$T_{\rm P,A=0,f=1}$}
\def\teqlday{$T_{\rm P,A=0,f=2}$}
\def\teqlinst{$T_{\rm P,A=0,f=8/3}$}

\begin{document}
   \title{Thermal emission from WASP-24b at 3.6 and 4.5 \micron}

\authorrunning{A. M. S. Smith \textit{et al.}}
\titlerunning{Thermal emission from WASP-24b}

\author{
	A. M. S. Smith\inst{\ref{inst1}}$^\dagger$\and
          D. R. Anderson\inst{\ref{inst1}}\and
          N. Madhusudhan\inst{\ref{inst2}}\and
          J. Southworth\inst{\ref{inst1}}\and
          A. Collier Cameron\inst{\ref{inst3}}\and
          J. Blecic\inst{\ref{inst4}}\and
          J. Harrington\inst{\ref{inst4}}\and
          C. Hellier\inst{\ref{inst1}}\and
          P. F. L. Maxted\inst{\ref{inst1}}\and
          D. Pollacco\inst{\ref{inst5}}\and
          D. Queloz\inst{\ref{inst6}}\and
          B. Smalley\inst{\ref{inst1}}\and
          A. H. M . J. Triaud\inst{\ref{inst6}}\and
    	P. J. Wheatley\inst{\ref{inst7}}      
                    }

\institute{
	Astrophysics Group, Lennard-Jones Laboratories, Keele University, Keele, Staffordshire, ST5 5BG, UK\label{inst1} \and
	Department of Physics, and Department of Astronomy, Yale University, New Haven, CT 06511, USA\label{inst2}\and
	SUPA, School of Physics and Astronomy, University of St. Andrews, North Haugh, St. Andrews, Fife, KY16 9SS, UK\label{inst3}\and
	Planetary Sciences Group, Department of Physics, University of Central Florida, Orlando, FL 32816-2385, USA\label{inst4}\and
	Astrophysics Research Centre, Physics Building, Queen's University, Belfast, County Antrim, BT7 1NN, UK\label{inst5}\and
	Observatoire de Gen\`{e}ve, Universit\'{e} de Gen\`{e}ve, 51 Chemin des Maillettes, 1290 Sauverny, Switzerland\label{inst6}\and
	Department of Physics, University of Warwick, Coventry CV4 7AL, UK\label{inst7}
	\\
              \newline$^\dagger$\email{a.m.s.smith@keele.ac.uk}
             }

   \date{Received March 23, 2012; accepted August 17, 2012}

 
  \abstract
   {}
   {We observe occultations of WASP-24b to measure brightness temperatures and to determine whether or not its atmosphere exhibits a thermal inversion (stratosphere).}
   {We observed occultations of WASP-24b at 3.6 and 4.5 \micron using the Spitzer Space Telescope. It has been suggested that there is a correlation between stellar activity and the presence of inversions, so we analysed existing HARPS spectra in order to calculate \rhk for WASP-24 and thus determine whether or not the star is chromospherically active. We also observed a transit of WASP-24b in the Str\"{o}mgren u and y bands, with the CAHA 2.2-m telescope.}
   {We measure occultation depths of $0.159 \pm 0.013$ per cent at 3.6~\micron and $0.202 \pm 0.018$ per cent at 4.5~\micron. The corresponding planetary brightness temperatures are $1974 \pm 71$~K and $1944 \pm 85$~K respectively. Atmosphere models with and without a thermal inversion fit the data equally well; we are unable to constrain the presence of an inversion without additional occultation measurements in the near-IR. We find \rhk~$= -4.98 \pm 0.12$, indicating that WASP-24 is not a chromospherically active star. Our global analysis of new and previously-published data has refined the system parameters, and we find no evidence that the orbit of WASP-24b is non-circular.}
   {These results emphasise the importance of complementing \spitz measurements with observations at shorter wavelengths to gain a full understanding of hot Jupiter atmospheres.}


   \maketitle

\section{Introduction}

Observing the occultation of an exoplanet by its host star at near-infrared wavelengths allows the direct detection of thermal emission from the planet. Since the first such measurement with the {\it Spitzer Space Telescope} \citep{Charbonneau-etal05,Deming05}, the occultations of numerous planets have been measured using both \spitz at 3.6 to 24 \micron, and ground-based telescopes at shorter wavelengths. 

Measuring the depth of a single occultation allows us to measure the brightness temperature of the dayside of the planet and estimate how efficiently heat is redistributed from the dayside to the nightside. Occultation measurements made at multiple wavelengths enable the construction of a spectral energy distribution (SED) of the planet's dayside atmosphere. The atmospheric depth probed by observations at a given wavelength depends on the opacity of the atmosphere in that passband, meaning that  a spectrum can inform us of the vertical structure of the atmosphere. In particular, thermal inversions (stratospheres), where the temperature increases with altitude, have been inferred for some planetary atmospheres on the basis of occultation data.

It was previously suggested that inversions are present in atmospheres hot enough to maintain high-opacity absorbers (such as TiO and VO) in the upper atmosphere \citep{hubeny03,burrows07,fortney}. More recent results \citep{Machalek08,Fressin10,madhu10} and theory \citep{Spiegel09}, however, have challenged this picture. An alternative hypothesis has been put forward by \cite{knutson_stellar_activity}, who note the apparent correlation between stellar activity and the absence of a thermal inversion. The planets that lack thermal inversions orbit chromospherically active stars; it is proposed that the high levels of ultraviolet flux associated with chromospheric activity destroy the high-altitude compounds that would otherwise cause inversions. This hypothesis is based on a small number of systems; observations of more planets are required to test it. Yet another possibility is that TiO and VO can be naturally low in abundance if the atmosphere is carbon-rich, C/O $\geq$ 1, and hence preclude thermal inversions \citep{madhu_carbon}.

WASP-24 is a late-F star which is host to a massive (1.1 \mjup) transiting planet in a short-period (2.34~d), prograde orbit \citep{w24,w24rm}. The stellar flux received by WASP-24b ($2.28\times10^9~\mathrm{Wm^{-2}}$) places the planet well into the pM class proposed by \cite{fortney}, so according to this theory a strong inversion should be present. In this paper, we present \spitz occultation observations of WASP-24b at 3.6 and 4.5~\micron and we measure \rhk for WASP-24.

\section{New observations}
\subsection{{\it Spitzer} occultation photometry}

We observed two occultations of the planet WASP-24b by its host star with the {\em Spitzer Space Telescope} \citep{spitzer}, during its warm mission, on 2011 April 16 and 2011 April 18. We used the Infrared Array Camera (IRAC; \citealt{IRAC}) in full-array mode ($256\times256$\,pixels, 1.2~\arcsec\ pixel$^{-1}$). The first occultation was observed at 3.6~\micron\ (channel 1) and the second at 4.5~\micron\ (channel 2). Each observation consisted of a total of 2054 10.4-s exposures, spanning a total of 7.46 hours.

The target was placed near the centre of the array for the channel 2 observations, but for the channel 1 observations it was placed on the pixel with $x$ and $y$ co-ordinates \{79,55\}. This position was chosen because \cite{Todorov10} found low intrapixel sensitivity variation during observations of HAT-P-1 centred at \{79.0,55.6\}. Our observations were conducted with the target star on a different part of the same pixel (centred on \{79.0,55.4\}), with no positional overlap with the \cite{Todorov10} observations. A strong position-dependent flux variation is observed in our data(see Sec. \ref{sec:analysis}).\\

We used the images calibrated by the standard {\it Spitzer} pipeline (version S18.18.0) and delivered to the community as Basic Calibrated Data (BCD). We performed the photometry using the method presented in \cite{wasp17_spitzer}. For each image, we converted flux from MJy~sr$^{-1}$ to electrons and then used {\sc iraf} to perform aperture photometry of the target, using circular apertures with radii ranging from 1.5 to 6.0 pixels. The apertures were centred by fitting 1D Gaussians with a fixed FWHM (1.37 and 1.62 pixels for channels 1 and 2 respectively) to the marginal profiles in $x$ and $y$, using a non-linear least squares technique. We measured the sky background in an annulus centered on the target with an inner radius of 8 pixels and an outer radius of 12 pixels, and subtracted it from the flux measured within the on-source apertures.

The flux in the sky background annulus was calculated using {\sc iraf}'s centroid algorithm, which is equivalent to calculating the mode of the histogram of sky background values. We note that warm Spitzer suffers from a systematic known as `column pulldown', where a column of pixels records anomalously high flux readings (see \citealt{deming11} for a discussion of systematic effects in warm Spitzer data). One such column pulldown associated with a hot pixel in the 4.5~\micron detector is located sufficiently close to our target that a it appears within our sky background annulus. This does not constitute a problem for our estimation of the sky background, however, since only a small fraction ($\approx$~15 / 250) of pixels are affected and sigma clipping is performed as part of the sky background estimation algorithm, so the affected pixels are rejected from the calculation.

We estimated the photometric uncertainty as the quadrature addition of the uncertainty in the sky background (estimated as the standard deviation of the flux in the sky anulus) in the on-source aperture, the read-out noise, and the Poisson noise of the total background-subtracted counts within the on-source aperture. The mid-exposure times (in the HJD (UTC) time system) are calculated by adding half the integration time (FRAMTIME value in the image headers) to the MHJD\_OBS values, which are the start times of the DCEs (Data Collective 
Events). These times were then converted to the BJD (TDB) system using the routines of \cite{Eastman_BJD}.

We adopt the aperture radii (2.6 pixels for channel 1 and 2.4 pixels for channel 2) that give the highest signal-to-noise, and the smallest residual scatter in the fitted light curves (see section \ref{sec:analysis} for details of our fitting procedure). The variation in both the channel 1 and 2 occultation depths is much smaller than 1~$\sigma$ across a wide range in aperture radius.

\begin{figure*}
\begin{center}
$\begin{array}{ccc}
\includegraphics[]{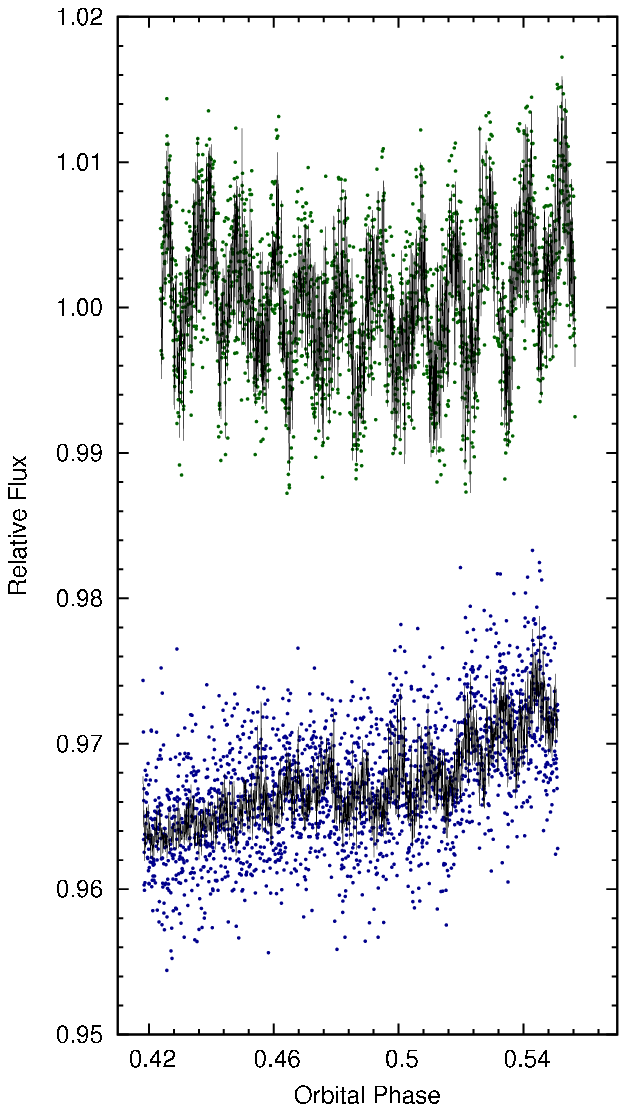} &
\includegraphics[]{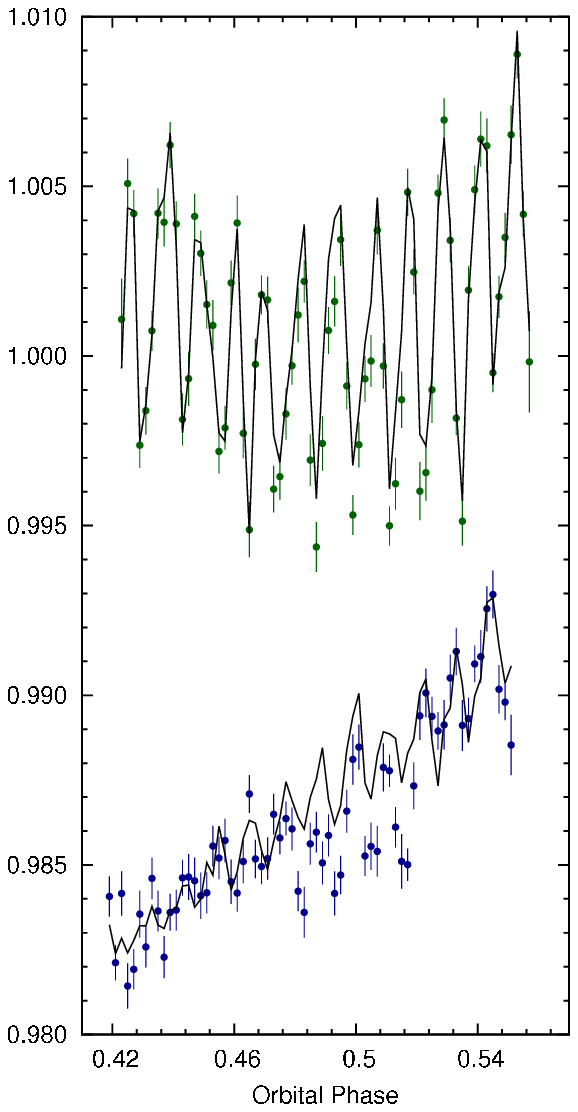} &
\includegraphics[]{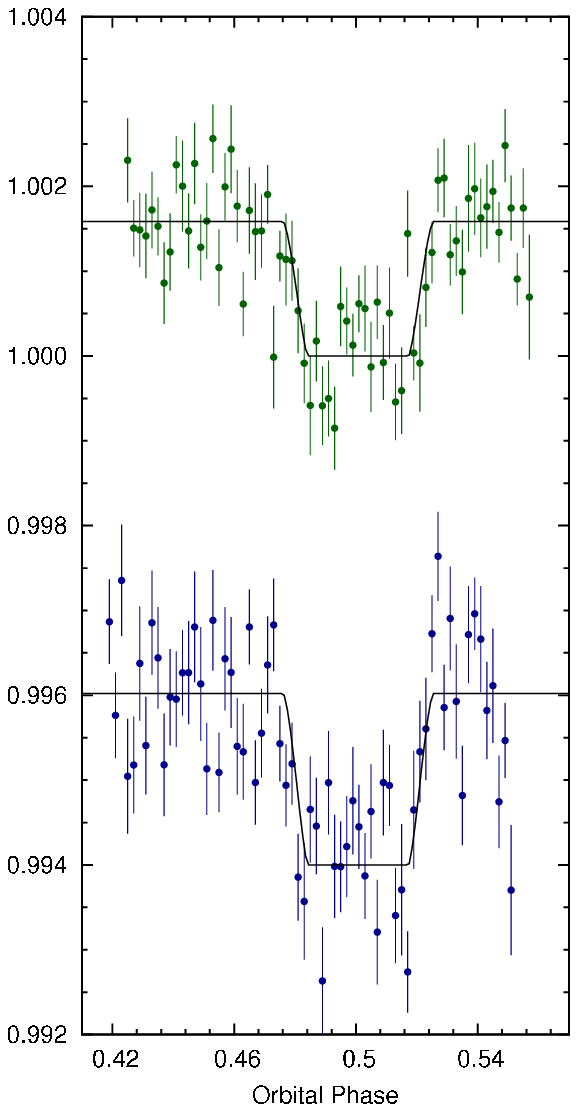}
\end{array}$
\caption{
In each of the above three plots, the 3.6 \micron data (green points) are shown above the 4.5 \micron data (blue points), which is offset in relative flux for clarity. 
{\bf\em Left}: Raw \spitz data with the best-fitting trend and 
occultation models superimposed.
{\bf\em Middle}: The same data binned in phase ($\Delta \phi=0.002 \approx 6.74$ 
min) with the best-fitting trend models superimposed.
{\bf\em Right}: The binned data after dividing by the best-fitting trend models, 
and with the best-fitting occultation models superimposed.
The error bar on each binned measurement in the panels in the middle and on the 
right is the standard deviation of the points within the bin.
\label{fig:spitz}}
\end{center}
\end{figure*}

In order to account for cosmic-ray hits, we rejected any flux measurement that was discrepant with the median of its 20 neighbours (a window width of 4.4 min) by more than four times its theoretical error bar. We also performed a rejection on target position. For each image and for the {\it x} and {\it y} detector coordinates separately, we computed the difference between the fitted target position and the median of its 20 neighbours. For each dataset, we then calculated the standard deviation, $\sigma$, of these {\it median differences} and rejected any points discrepant by more than 4 $\sigma$. We reject a total of 82 (4.0 per cent) channel 1 data points (73 on flux, 8 on x-position, and 1 on y-position) and 23 (1.1 per cent) channel 2 data points (18 on flux, 1 on x-position, and 4 on y-position). The post-rejection data are displayed raw and binned in the first and second panels respectively of Figure~\ref{fig:spitz}. 

\subsection{CAHA 2.2-m transit photometry}

\begin{figure}
\centering
\includegraphics[width=9cm]{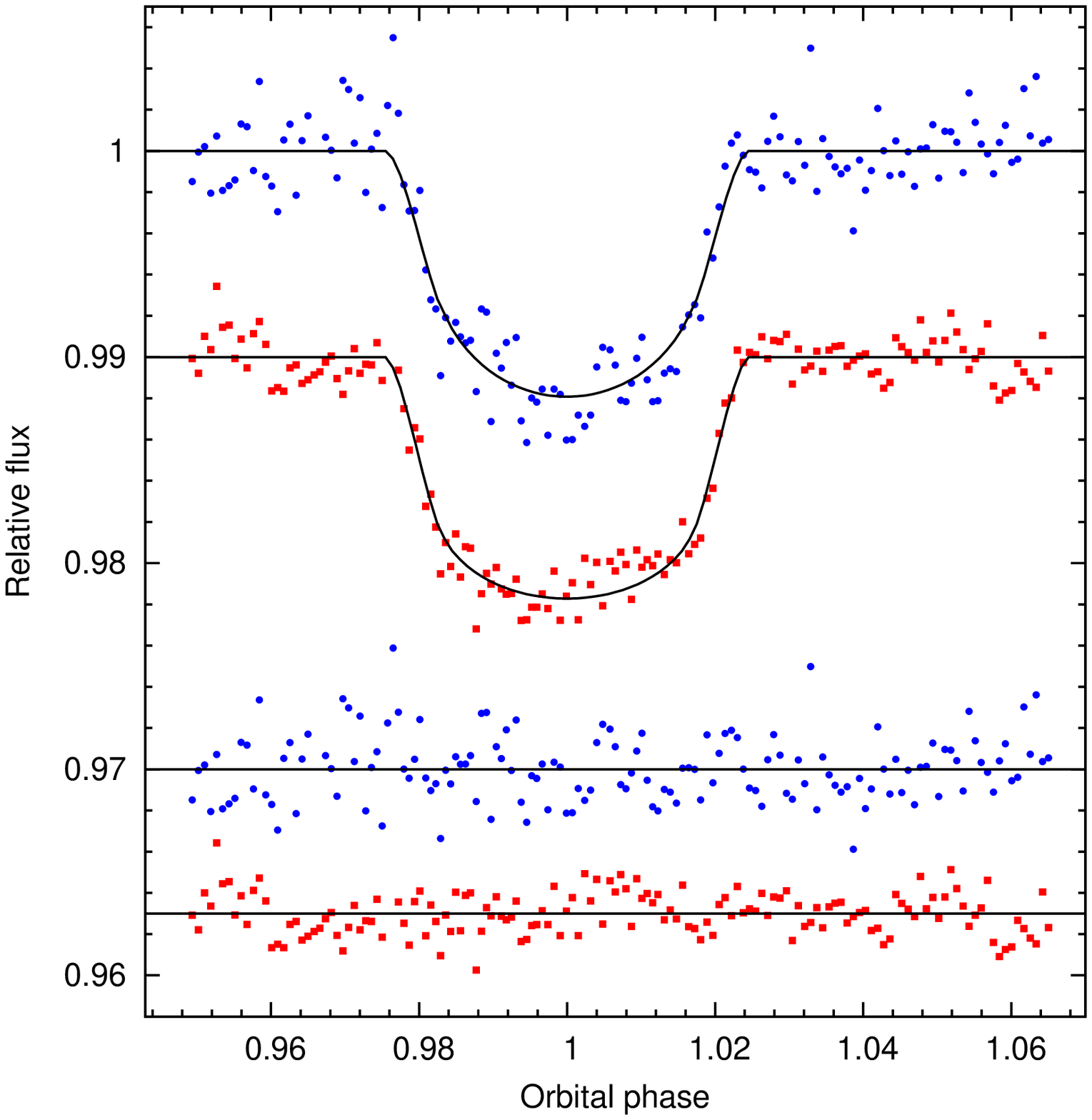}
\caption{Transit photometry from the CAHA 2.2-m telescope. The upper curve (blue circles) is the transit in the Str\"{o}mgren u passband, and the curve below it (red squares) is the transit in Str\"{o}mgren y, offset in flux for clarity. In each case our best fitting model is overplotted (solid line) and the residuals are shown beneath, offset from zero and with the same symbols as the corresponding light curve.
}
\label{fig:busca}
\end{figure}

We observed a transit of WASP-24b on 2010 May 05/06 with the CAHA 2.2-m telescope at Calar Alto, Spain. We used the Bonn University Simultaneous CAmera (BUSCA) to observe simultaneously in the Str\"{o}mgren u, b and y passbands, and the SDSS z passband, with exposure times of 90 to 120 s. Further details of our BUSCA observation strategy may be found in \cite{jkt_hat5}. Aperture photometry was performed using the pipeline of \cite{Southworth-wasp5}, which uses the {\sc idl} implementation of the {\sc daophot aper} algorithm \citep{daophot} provided by the {\sc astrolib} library. An ensemble of comparison stars were used and a systematic trend in the form of a linear function of time was removed. The b and z-band data were both found to exhibit an unacceptably large scatter, the cause of which is unknown; we use only the Str\"{o}mgren u and y data in our subsequent analysis. These data are shown overplotted with our best-fitting model in Figure \ref{fig:busca}.

\section{Data analysis}
\label{sec:analysis}

Our new \spitz data and transit photometry were incorporated into a global analysis alongside the following data: a total of 29,993 photometric data points from SuperWASP-N and WASP-South, spanning 2008 March 5 to 2010 July 26 (including the 9750 presented by \citealt{w24} who included only the data taken prior to 2009 April 28); the three complete follow-up transit light curves of \cite{w24} from the Liverpool (V+R band), Faulkes North and Faulkes South Telescopes (both PanSTARRS-z band); the ten radial velocities (RVs) measured by the Nordic Optical Telescope, and the eighteen CORALIE RVs presented by \cite{w24}; and the 63 HARPS radial velocities of \cite{w24rm}. We analysed the data using an adaptive Markov-chain Monte Carlo (MCMC) algorithm (\citealt{Cameron-etal07}; \citealt{wasp3}; \citealt{Enoch10}; see \citealt{DRA_rm} for a description of the current version of our code).

The data that we include in our analysis from \cite{w24} and \cite{w24rm} were originally in the HJD (UTC) and BJD (UTC) time systems. We converted these to the BJD (TDB) system, using the routines described by \cite{Eastman_BJD}, so that there is consistency between all our data. In order to account for the light-travel time across the orbit when comparing transit and occultation photometry, we made a first-order adjustment to the timings of the \spitz data, subtracting 36 s from the time of each observation (for comparison, we measure the mid-occultation time with a precision of 111 s).

The MCMC proposal parameters (which are defined in Table \ref{tab:sys-params}) we used are: $t_{\rm c}$, $P$, 
$(\rplanet/\rstar)^2$, $t_{14}$, $b$, $K_{\rm 1}$, \teff, \feh, \secos, \sesin, 
\svcos, \svsin, \iracone, and \iractwo. At each step in the MCMC procedure, each proposal parameter is perturbed from its previous value by a small, random amount. From the proposal parameters, model light and RV curves are generated and $\chi^{2}$ is calculated from their comparison with the data. A step is accepted if $\chi^{2}$ (our merit function) is lower than for the previous step, and a step with higher $\chi^{2}$ is accepted with probability $\exp(-\Delta \chi^{2}/2)$. In this way, the parameter space around the optimum solution is thoroughly explored. The value and uncertainty for each parameter are taken as the median and central 68.3 per cent confidence interval of the parameter's marginalised posterior probability distribution respectively \citep[e.g.][]{Ford06}.

The transit light curves were modelled using the formulation of \cite{M&A} and limb darkening was accounted for using a four-coefficient, non-linear model, employing coefficients appropriate to the passband from the tabulations of \cite{Claret, Claret04}. The coefficients were determined using an initial interpolation in $\log g_{*}$ and [Fe/H] (using the values reported in \citealt{w24}), and an interpolation in $T_{\rm *, eff}$  at each MCMC step. The coefficient values corresponding to the best-fitting value of $T_{\rm *, eff}$ are given in Table \ref{tab:limb}.

\subsection{De-trending Spitzer data}
\label{sec:trend}

Both the IRAC 3.6 and 4.5 \micron~ detectors are known to have an inhomogeneous intra-pixel sensitivity which results in a correlation between the measured flux of a star and its position on the array \citep[e.g.][and references therein]{Knutson_hd209}. This, combined with a small, periodic ($P \approx 40$min) pointing wobble thought to be caused by thermal cycling of a heater aboard the spacecraft\footnote{http://ssc.spitzer.caltech.edu/warmmission/news/21oct2010memo.pdf}, results in a periodic modulation of the photometry (Fig. \ref{fig:spitz}). We model this so-called `pixel-phase' effect as a quadratic function of the $x$ and $y$ positions of the centre of the PSF, after \cite{charbonneau_hd189}. We include in our model the additional cross-term of \cite{desert_hd189} as well as a term linear in time, such that the stellar flux relative to its weighted mean, $df = f - \hat{f}$, is given by

\begin{equation}
\label{eqn:ch2}
df = a_0 + a_xdx + a_ydy + a_{xy}dxdy + a_{xx}dx^2 + a_{yy}dy^2 + a_tdt
\end{equation}

\noindent where $dx = x - \hat{x}$ and $dy = y - \hat{y}$ are the coordinates of the PSF 
centre relative to their weighted means, $dt$ is the time elapsed since the 
first observation, and $a_0$, $a_x$, $a_y$, $a_{xy}$, $a_{xx}$, 
$a_{yy}$ and $a_t$ are the coefficients for which we fit.

At each MCMC step, after dividing the data by the eclipse model we determine the trend model coefficients using the singular value decomposition routine of \cite{numrec}. The best-fitting trend models are binned and superimposed on the binned photometry in the middle panel of Figure~\ref{fig:spitz}, and the trend model coefficients are given in Table \ref{tab:coeffs}.

\subsection{Photometric noise}

\begin{figure}
\centering
\includegraphics[width=9cm]{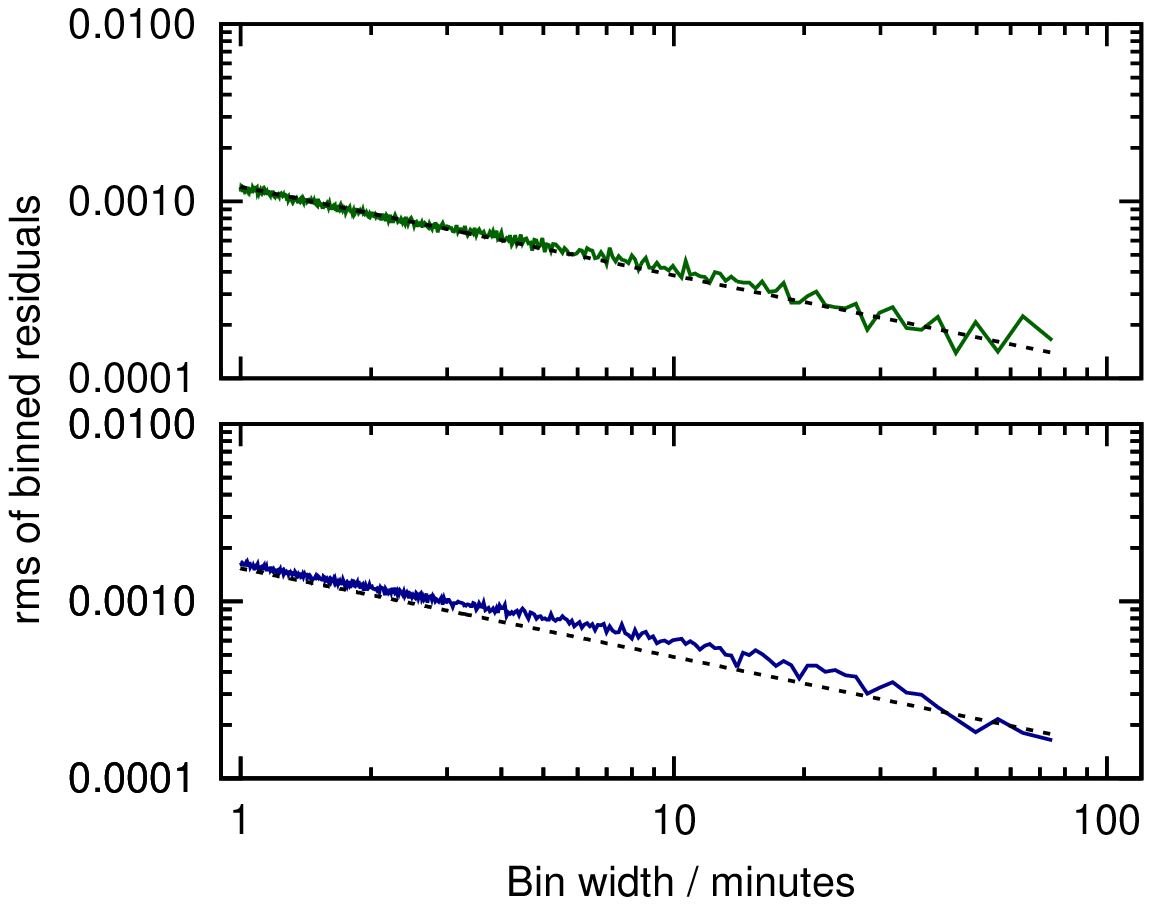}
\caption{The rms of the binned residuals for the \spitz data at 3.6 \micron (upper) and 4.5 \micron (lower) (solid lines). The white-noise expectation, where the rms decreases in proportion to the square root of the bin size, is indicated by the dashed line in each panel.
}
\label{fig:rms}
\end{figure}

We scaled the formal uncertainties of each photometric dataset by an individual factor in order to obtain a reduced $\chi^2$ of unity. This ensures that each dataset is properly weighted with respect to the others in the MCMC analysis and that realistic uncertainties are obtained. The errors on the transit photometry were multiplied by factors ranging from 0.95 to 2.62, while those on the \spitz channel 1 and 2 data were multiplied by 1.070 and 1.007 respectively. It was not necessary to add any jitter to the RV uncertainties in order to obtain a reduced spectroscopic-$\chi^2$ of unity. We plotted the rms of the binned residuals as a function of bin width to assess the presence of correlated noise in the \spitz light curves (Fig. \ref{fig:rms}). A small amount of correlated noise is detected in the 4.5\micron light curve on timescales of the order of 10 minutes.

Remaining correlated noise present in the occultation light curve can result in an underestimation of the uncertainty on the occultation depth. To check that our occultation depth uncertainties are not unrealistic, we employ the residual permutation or `prayer bead' method (e.g. \citealt{Gillon07}) on each of our \spitz light curves. We implement this method in the same way as we did in \cite{smith_w33}, i.e. we shift the residual to the model to the subsequent observation before adding back the trend and occultation models. This process is repeated, shifting each residual two observations, and so on, resulting in a series of light curves, each of which has the time-structure of the correlated noise preserved. A separate MCMC analysis was performed on each of these light curves. In order to expedite the process of conducting a total of 4007 analyses (1977 for channel 1, and 2030 for channel 2), we did not fit any of the transit photometry. Instead, we fixed the values of $t_{\rm c}$, $P$, $(\rplanet/\rstar)^2$, $t_{14}$ and $b$ to the values determined by our main MCMC analysis (the values listed in Table \ref{tab:sys-params}). We also fixed the values of $K_{\rm 1}$, \teff, \feh, \svcos, and \svsin, so that the only proposal parameters were those influenced by the occultation photometry, namely the occultation depth, \iracone~or \iractwo, and \secos~and \sesin.

We then calculated the median of the best-fitting occultation depths from the ensemble of new fits, as well as the uncertainty limits which enclose 68.3 per cent of the values around the median. The occultation depths from this prayer bead analysis are \iracone $= 0.159^{+ 0.018}_{- 0.010}$ per cent and \iractwo $= 0.210^{+ 0.016}_{- 0.024}$ per cent. These values and uncertainties are almost identical to those obtained from our global MCMC analysis, indicating that any residual correlated noise in our light curves has a negligible impact on our measurement of the occultation depths.

\begin{table}
\centering
\caption{Limb-darkening coefficients}
\label{tab:limb}
\begin{tabular}{cccccc} \hline 
Claret band & Light curves & $a_1$ &  $a_2$  &  $a_3$ & $a_4$  \\
\hline 
Cousins $R$  & WASP, RISE             & 0.581 &  0.011  & 0.373  &-0.236 \\
Sloan $z^\prime$        & FTS, FTN     & 0.673 & -0.353  & 0.600  &-0.303 \\
Str\"{o}mgren $u$ & BUSCA 	                   & 0.429 & -0.049  & 1.055  &-0.528 \\
Str\"{o}mgren $y$ & BUSCA 	                   & 0.498 &  0.224  & 0.277  &-0.206 \\
\hline 
\end{tabular} 
\end{table} 

\begin{table}
\caption{Trend model parameters and coefficients} 
\label{tab:coeffs} 
\centering
\begin{tabular}{l c c} 
\hline 
 		& 3.6 \micron			& 4.5 \micron		\\
\hline
$\hat{f}$	& 193116.39	        	& 103427.12	  \\
$\hat{x}$	& 79.03	                	& 127.44		  \\
$\hat{y}$	& 55.37 	        	& 129.41		  \\
$a_0$		& $ 63.9 \pm  5$        	& $ -84.8  \pm 2.4 $		\\
$a_x$		& $ 20586 \pm  18$      	& $ -756.4 \pm 5 $  	\\
$a_y$		& $ -21433 \pm  53$     	& $ -3937.5 \pm 6 $ 	\\
$a_{xy}$	& $ -9277 \pm  756$     	& $ -6791 \pm 329 $ 	\\
$a_{xx}$	& $ -26000 \pm  1302$   	& $ 7784 \pm 150 $	\\
$a_{yy}$	& $ 37604 \pm  891$     	& $ 8236 \pm 282 $ 	\\
$a_t$		& $ -633\pm  39$        	& $ 59 \pm 21 $		\\
\hline
\multicolumn{3}{l}{See Sec \ref{sec:trend} for the definitions of these parameters.}
\end{tabular}
\end{table}

\section{Results}
\label{sec:results}

\begin{table*} 
\caption{WASP-24 system parameters} 
\label{tab:sys-params} 
\begin{tabular}{llcl}
\hline 
Parameter & Symbol & Value & Unit\\ 
\hline 
\\
Orbital period & $P$ & $ 2.3412132\pm 0.0000018$ & d\\
Epoch of mid-transit (BJD, TDB) & $t_{\rm c}$ & $2455149.27535 \pm 0.00015$ & d\\
Transit duration (from first to fourth contact) & $t_{\rm 14}$ & $0.11364 \pm 0.00069$ & d\\
Duration of transit ingress $\approx$ duration of transit egress & $t_{\rm 12} \approx t_{\rm 34}$ & $0.01839 \pm 0.00082$ & d\\
Planet-to-star area ratio & (\rplanet/\rstar)$^2$ & $0.01102 \pm 0.00013$ & \\
Impact parameter & $b$ & $0.669 \pm 0.016$ & \\
Orbital inclination & $i$ & $83.30 \pm 0.30$ & $^\circ$\\
\noalign{\medskip}
Semi-amplitude of the stellar reflex velocity & $K_{\rm 1}$ & $150.6 \pm 2.5$ & \ms\\
Centre-of-mass velocity & $\gamma_{\rm rv1}$ & $-17803 \pm 0.15$ & \ms\\
Offset between CORALIE and NOT & $\gamma_{\rm CORALIE-NOT}$ & $102.85 \pm 0.15$ & \ms\\
Offset between CORALIE and HARPS & $\gamma_{\rm HARPS-CORALIE}$ & $-14.21 \pm 0.64$ & \ms\\
\noalign{\medskip}
Orbital eccentricity & $e$ & $0.0031^{+ 0.0097}_{- 0.0024}$ & \\
& & $< 0.0388$ (3 $\sigma$) & \\
Argument of periastron & $\omega$ & $67^{+ 24}_{- 149}$ & $^\circ$\\
& $e\cos\omega$ & $0.00044^{+ 0.00097}_{- 0.00076}$ & \\
& $e\sin\omega$ & $0.0009^{+ 0.0011}_{- 0.0028}$ & \\
Phase of mid-occultation, having accounted for light travel time & $\phi_{\rm mid-occ.}$ & $0.50028^{+ 0.00062}_{- 0.00049}$ & \\
Occultation duration & $t_{\rm 58}$ & $0.11395^{+0.00101}_{-0.00088}$ & d\\
Duration of occultation ingress $\approx$ duration of occultation egress & $t_{\rm 56} \approx t_{\rm 78}$ \medskip & $0.01862^{+ 0.00104}_{- 0.00091}$ & d\\
\noalign{\medskip}
Relative planet-to-star flux at 3.6  \micron & \iracone	& $0.159 \pm 0.013$ & per cent\\
Relative planet-to-star flux at 4.5  \micron & \iractwo	& $0.202 \pm 0.018$ & per cent\\
Planet brightness temperature at 3.6 \micron & $T_{\rm b,3.6}$ & $1974 \pm 71$ & K\\
Planet brightness temperature at 4.5 \micron & $T_{\rm b,4.5}$ & $1944 \pm 85$ & K\\
\noalign{\medskip}
Sky-projected stellar rotation velocity & \vsini & $5.86 \pm 0.63$ & \kms\\
Sky-projected angle between stellar spin and planetary orbit axes & $\lambda$ & $-5.8 \pm 4.1$ & $^\circ$\\
\noalign{\medskip}
Stellar mass & \mstar & $1.154 \pm 0.025$ & \msol\\
Stellar radius & \rstar & $1.354 \pm 0.032$ & \rsol\\
Stellar density & \densstar & $0.465 \pm 0.028$ & \denssol\\
Stellar surface gravity & $\log g_{*}$ & $4.236 \pm 0.017$ & (cgs)\\
Stellar effective temperature & $T_{\rm *,eff}$ & $6038 \pm 95$ & K\\
Stellar metallicity & \feh & $0.099 \pm 0.071$ & (dex)\\
\noalign{\medskip}
Planet mass & \mplanet & $1.091 \pm 0.025$ & \mjup\\
Planet radius & \rplanet & $1.383 \pm 0.039$ & \rjup\\
Planet density & \densplanet & $0.412 \pm 0.032$ & \densjup\\
Planet surface gravity & $\log g_{\rm P}$ & $3.115 \pm 0.022$ & (cgs)\\
Scaled orbital major semi-axis & $ a/\rstar$ & $5.75 \pm 0.11 $ & \\
Orbital major semi-axis & $a$ & $0.03619 \pm 0.00027$ & AU\\
Planet equilibrium temperature$^{\dagger}$ (full redistribution) & \teqlfull & $ 1781 \pm 34 $ & K\\
Planet equilibrium temperature$^{\dagger}$ (day side redistribution) & \teqlday & $ 2118 \pm 41 $ & K\\
Planet equilibrium temperature$^{\dagger}$ (instant re-radiation) & \teqlinst & $ 2276 \pm 44 $ & K\\
\\ 
\hline 
\multicolumn{4}{l}{$^{\dagger}$ 
\teqlf\ $= f^{\frac{1}{4}}T_{\rm eff}\sqrt{\frac{R_*}{2a}}$ where $f$ is the 
redistribution factor, with $f=1$ for full redistribution, $f=2$ for dayside 
redistribution}\\
\multicolumn{4}{l}{and $f=8/3$ for instant re-radiation 
\citep{CowanAgol}. We assumed the planet albedo to be zero, $A=0$.}
\end{tabular} 
\end{table*} 

\subsection{Brightness temperatures}
\label{sec:bright}

The median values of the system parameters and their associated $1\sigma$ uncertainties from our MCMC analysis are displayed in Table \ref{tab:sys-params}. Our occultation models are shown with the de-trended \spitz photometry in the third panel of Figure \ref{fig:spitz}. We find occultation depths of $0.159 \pm 0.013$ percent at 3.6~\micron and $0.202 \pm 0.018$ per cent at 4.5~\micron. We calculated corresponding brightness temperatures by equating the occultation depth with the product of (\rplanet/\rstar)$^2$ and the ratio of the planet-to-star flux density (calculated using a model stellar spectrum of \citealt{kurucz}) integrated over the bandpass and corrected with the \spitz transmission curve. We then scaled the planetary blackbody temperature so that the occultation depth equalled the calculated quantity. We find brightness temperatures of $1974 \pm 71$~K at 3.6~\micron and $1944 \pm 85$~K at 4.5~\micron. These uncertainties account for only the uncertainty in the occultation depth, which is the dominant source of error.

\subsection{Orbital eccentricity}

\cite{w24} found no evidence for a non-zero orbital eccentricity of WASP-24b, and adopted a circular orbital solution. Given the additional constraints that occultation observations place on \esin, and particularly \ecos, we chose not to constrain the orbital eccentricity within our MCMC analysis. We do not detect a non-zero eccentricity with any significance, however. Our median value is $e = 0.0031^{+ 0.0097}_{- 0.0024}$, and our $3\sigma$ upper limit to the eccentricity is 0.0388. Our best-fitting phase of mid-occultation, $\phi_{\rm mid-occ.} = 0.50028^{+ 0.00062}_{- 0.00049}$, corresponds to an occultation occurring $57^{+ 125}_{- 99}$~s later than expected for a circular orbit.

\subsection{System parameters}

The system parameters resulting from our analysis should be viewed as superseding previous measurements by \cite{w24} and \cite{w24rm}, since we include in our analysis all of the data analysed by those authors as well as additional, new data. We made a comparison of the parameters presented in Table \ref{tab:sys-params} with the corresponding values (where given) from \cite{w24} and \cite{w24rm}, finding them to be in very good agreement. The majority of our measurements differ from previously published values by less than 1-$\sigma$; in no case is the deviation greater than 2-$\sigma$.

\section{Discussion}

\subsection{Brightness and equilibrium temperatures}

The fact that the 3.6 and 4.5~\micron brightness temperatures are essentially identical (Table \ref{tab:sys-params}; Section \ref{sec:bright}) is suggestive that no strong temperature inversion exists in the atmosphere of WASP-24b. The observed occultation depths are well fitted by a blackbody SED for the planet (Fig. \ref{fig:model}), with a best-fitting blackbody temperature of 1952 $\pm$ 52~K.

The brightness temperatures may be compared with equilibrium temperatures calculated under various assumptions. The equilibrium temperature for a zero-albedo planet is given by
\begin{equation}
T_{\rm P, A=0} =  f^{\frac{1}{4}} ~T_{*,\rm eff}~\sqrt{\frac{R_*}{2a}},
\end{equation}
where $f$ parameterises the redistribution of heat; $f = 1$ indicates isotropic re-radiation over the whole planet (i.e. the redistribution of heat from the dayside to the nightside is fully efficient). The case where the heat is uniformly distributed on the dayside of the planet, but there is no redistribution of heat to the nightside  corresponds to $f = 2$. A third case, corresponding to $f = \frac{8}{3}$ (equivalent to $\varepsilon = 0$ and $f = \frac{2}{3}$ in the notations of \citealt{CowanAgol} and \citealt{LM_S} respectively), occurs when the incident radiation is immediately re-radiated and there is no redistribution of heat within even the dayside. Because the hottest region of the dayside is most visible close to occultation, a deeper occultation is observed than if the dayside had a uniform temperature.

The zero-albedo equilibrium temperatures for the three aforementioned cases are presented at the bottom of Table \ref{tab:sys-params}. Our brightness temperatures fall between the equilibrium temperatures for $f = 1$ and $f = 2$, suggesting that there is significant redistribution of heat to the nightside. We find a dimensionless dayside effective temperature, the $T_\mathrm{d}/T_0$ of \cite{CowanAgol} (where $T_\mathrm{d}$ is the dayside effective temperature, and $T_0$ is the equilibrium temperature of the sub-stellar point), of $0.77 \pm 0.03$. In a plot of $T_\mathrm{d}/T_0$ vs. the maximum-expected dayside temperature (Fig. 7 of \citealt{CowanAgol}), WASP-24b is placed close to TrES-4b, amongst the main cluster of points.

\subsection{Atmosphere model}

We model the dayside thermal emission from WASP-24b using the exoplanetary atmospheric modeling and retrieval method of \cite{Madhu09,madhu10}. The model computes line-by-line radiative transfer in a plane parallel atmosphere in local thermodynamic equilibrium. The model assumes hydrostatic equilibrium and global energy balance. The model atmosphere includes the major sources of opacity expected in hot hydrogen-dominated giant planet atmospheres, namely, molecular absorption due to H$_2$O, CO, CH$_4$, and CO$2$, and continuum opacity due to H$_2$-H$_2$ collision-induced absorption (CIA). Our molecular line-lists were obtained from \cite{Freedman08}, Freedman (2009, private comm.), and \cite{Rothman05}, and we use the CIA opacities from \cite{Borysow97} and \cite{Borysow02}. The chemical composition and the temperature structure of the atmosphere are input parameters to the model. Given a set of atmospheric observations, we explore the space of chemical composition and temperature structure to determine the regions constrained by the data for a desired level of confidence (e.g. \citealt{Madhu_w12}). However, given that we have only two data points at present, a unique fit to the composition and temperature structure is not feasible. Consequently, we fix the molecular mixing ratios under the assumption of chemical equilibrium with solar elemental abundances (e.g. \citealt{Burrows99}), and explore the range of temperature structures required to explain the data.

\begin{figure}
\centering
\includegraphics[width=8cm]{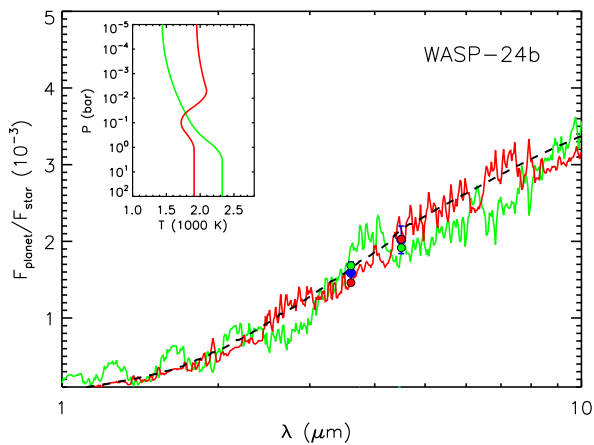}
\caption{Spectral energy distribution of WASP-24b relative to that of its host star. The blue circles with error bars are our best-fitting occultation depths. The light green line is a model-atmosphere spectrum, based on a model which assumes solar abundances in thermochemical equilibrium and lacks a temperature inversion, and the dark red line is a model with a temperature inversion. The band-integrated model fluxes are indicated with circles of the corresponding colours. The dashed black line shows a planetary black body model with a temperature of 1900 K. Inset: temperature-pressure profiles for our models. 
}
\label{fig:model}
\end{figure}

We find that both our planet-star flux ratios can be explained by a planetary blackbody at around 1950 K. Consequently, the data are consistent with an isothermal atmosphere. However, an isothermal temperature structure over the entire atmosphere is unrealistic in radiatively efficient atmospheres at low optical depth (e.g. \citealt{Hansen08}). More realistic is an atmosphere with temperature decreasing outward, i.e. one without a thermal inversion; such an atmosphere explains the data just as well (Fig. \ref{fig:model}, green curve). However, a second model with identical chemical composition, but with an inversion is an equally good fit to the data (Fig \ref{fig:model}, red curve). Both models predict very efficient day-night heat redistribution; the inversion model has 50 per cent redistribution, and the non-inversion model 40 per cent. Our two \spitz data points are therefore insufficient to constrain the presence of a thermal inversion in the atmosphere of WASP-24b.  The two models predict different flux ratios in the near-IR, so a measurement of the occultation depth at one or more near-IR wavelengths (e.g. J, H, K-bands) would break the degeneracy between the models.

\subsection{Thermal inversion}

\cite{knutson_stellar_activity} note a correlation between the presence / absence of thermal inversions in the atmospheres of hot Jupiters and stellar activity. Planets orbiting active stars are found to lack stratospheres, perhaps because the high-levels of ultraviolet flux associated with chromospherically active stars destroys the high-altitude sources of opacity thought to be responsible for thermal inversions. 

We determined the \rhk activity index \citep{noyes84} of WASP-24 by measuring the weak emission in the cores of the Ca {\sc ii} H+K lines, using the HARPS spectra obtained by \cite{w24rm} and the method of \cite{log_rhk}. Using the spectra with the highest signal-to-noise, we find \rhk = $-4.98 \pm 0.12$. This places (although not conclusively) WASP-24 among the inactive stars ($-5.5 < $\rhk$ < -4.9$) that \cite{knutson_stellar_activity} finds are host to planets with inversions. Furthermore, we find no evidence in the long-baseline WASP light curves for evidence of any rotational variability (there is no periodic variation above 1~mmag at a confidence level of 95 per cent). Rotational modulation is indicative of stellar chromospheric activity, and is observed in e.g. WASP-19, which has \rhk in the region of $-4.5$ \citep{w19_spitz}.

We calculate the model-independent, empirical metric of \cite{knutson_stellar_activity} for classifying hot Jupiter spectra, finding $\zeta = -0.005 \pm 0.025$ per cent \micron$^{-1}$ (adopting the notation of \citealt{w19_spitz}). This value suggests an inversion is present, but is close to the boundary between inverted and non-inverted atmospheres ($\approx -0.05$ per cent \micron$^{-1}$). On a plot of \rhk vs. $\zeta$ (Fig. 5 of \citealt{knutson_stellar_activity}), WASP-24b lies in the upper left corner of the cluster of planets classified as having inversions, very close to XO-1b.

\section{Conclusions}

We have detected thermal emission from WASP-24b at 3.6 and 4.5~\micron. Our measured occultation depths correspond to planetary brightness temperatures which are very similar in the two bandpasses; a blackbody planetary model is a good fit to the data. Theoretical model atmospheres which have solar abundances, both with and without a thermal inversion are also a good fit (Fig. \ref{fig:model}). We are unable to constrain the presence of the thermal inversion in the atmosphere predicted by both the TiO/VO \citep{fortney} and the stellar activity \citep{knutson_stellar_activity} hypotheses.

With \rhk~$=4.98 \pm 0.12$, WASP-24 appears to be chromospherically inactive, however the relatively large uncertainty on this value means that the star is close to the boundary between active and inactive stars, although nothing else about the star suggests that it has an active chromosphere. This system would therefore present a challenge to the stellar activity hypothesis if near-IR occultation measurements were found to indicate that the atmosphere lacks an inversion. 

We find the orbit of WASP-24b to be consistent with a circular orbit; we place a $3\sigma$ upper limit of 0.039 to the orbital eccentricity.

\begin{acknowledgements}
This work is based on observations made with the Spitzer Space Telescope, which is operated by the Jet Propulsion Laboratory, California Institute of Technology under a contract with NASA. JH and JB acknowledge NASA support through an award issued by JPL/Caltech.
\end{acknowledgements}

\bibliographystyle{aa}
\bibliography{iau_journals,refs2}

\end{document}